\input amstex
\documentstyle{amsppt}
\def\id{\text{id}\,}

\def\CC{\text{\bf C}}
\def\rank{\text{rank\,}}
\def\hom{\text{hom\,}}
\def\ev{\text{ev\,}}
\def\sq{\square}
\topmatter
\title Representation theory, topological field theory, and the
Andrews-Curtis conjecture\endtitle
\rightheadtext{Representations and the Andrews-Curtis conjecture}
\author Frank Quinn\endauthor
\address Virginia Tech, Blacksburg VA 24061-0123\endaddress
\email Quinn@vtmath.math.vt.edu\endemail
\date February 1992\enddate

\abstract
We pose a representation-theoretic question motivated by an
attempt to resolve the Andrews-Curtis conjecture. Roughly, is there a
triangular Hopf algebra with a collection of self-dual irreducible
representations $V_i$ so that the product of any two decomposes as
a sum of copies of the $V_i$, and $\sum (\rank V_i)^2=0$? This data
can be used to construct a ``topological quantum field theory'' on 2-
complexes which stands a good chance of detecting counterexamples
to the conjecture.
\endabstract
\endtopmatter

The first section recalls the Andrews-Curtis conjecture and its
analogy with diffeomorphisms of 4-manifolds. The second section
suggests the relevance of topological field theory. In the third we
precisely state the representation-theoretic question.   The final
section sketches how the data is used to construct a field theory on
2-complexes.

\head 1. The Andrews-Curtis conjecture\endhead
The conjecture asserts that if two 2-dimensional CW complexes are
simple homotopy equivalent then there is a deformation from one to
the other through 2-complexes. In a deformation we allow
expansions and collapses of cells of dimension $\le2$, and homotopy
of the attaching maps of 2-cells. There is a combinatorial group
theory formulation obtained by thinking of 2-complexes as
presentations of groups.

The special case of contractible complexes is particularly provocative:
can a contractible 2-complex be deformed to a point? As specific
examples we note the presentations
$$\langle x,y\mid xyx=yxy, x^m=y^n\rangle$$
determine 2-complexes which are contractible if $m=n\pm1$. This
can be deformed to a point if $\{m,n\}=\{2,3\}$ (S. Gersten), but is
expected to be a counterexample for other values \cite{AK}, \cite{G}.

The conjecture in the contractible case arose in an approach to the 4-
dimensional smooth Poincar\'e conjecture \cite{AC}. There is some
literature on the question, eg\. \cite{M}, \cite{MH}, but this
considerably underrepresents the attention it has received because
there has been almost no progress. If true the conjecture would have
useful applications in the study of low dimensional smooth manifolds
\cite{AK}, \cite{Q1}, \cite{G}, \cite{C2}. If it is false as expected there
do  not seem to be direct consequences for manifolds \cite{C1}. In
this case its greatest attraction may be as a model problem for
invariants of smooth 4-manifolds, as we explain next.

Exotic smooth structures on 4-manifolds have a strong nilpotence
property: Suppose $f\:M\to N$ is a map which the high-dimensional
theory predicts should be homotopic to a diffeomorphism (see eg\.
\cite{FQ, Ch. 7}). Then for some $k$ the map $f\#\id\:M\#kS^2\times
S^2\to N\#kS^2\times S^2$ is homotopic to a diffeomorphism
\cite{Q2}. For the usual sorts of algebraic-topological invariants one
can recover invariants of $f$ from those of $f\#\id$, so these
invariants cannot detect exotic $f$. Donaldson \cite{DK} has shown
exotic maps exist, using invariants defined with the space of
solutions to anti-self-dual Yang Mills equations. These invariants are
killed by connected sum with $S^2\times S^2$. The proper context for
these invariants is far from clear. Also they leave the smooth
Poincar\'e conjecture untouched. It would be very helpful to have
examples of similar invariants to study for clues on how to deal with
these problems.

The Andrews-Curtis problem has a very similar nilpotence property:
Suppose $f\:X\to Y$ is a simple homotopy equivalence of 2-
complexes. Then for some $k$ the map $f\vee \id\:X\vee kS^2\to
Y\vee kS^2$ is homotopic to a deformation through 2-complexes.
Here $X\vee kS^2$ indicates the 1-point union with $k$ 2-spheres.
Again traditional invariants of $f$ can be recovered from those of
$f\vee\id$, so cannot detect counterexamples to the conjecture.
Because of this similarity we expect invariants capable of detecting
exotic 2-complexes to behave much like 4-manifold invariants. They
might also be simpler, displaying algebra and topology without the
heavy burden of analysis and geometry.

\head 2. Topological field theory\endhead
Atiyah \cite{A} has suggested that the Donaldson invariants might be
some sort of ``topological quantum field theory'' (TQFT). Floer,
Donaldson, and others have partly implemented this intuition, though
profound mysteries remain.

Atiyah also suggested there might be interesting TQFT on 3-
manifolds. Witten \cite{W1} gave a much more detailed prediction
which has been implemented in several cases (\cite{Wa}, \cite{KM},
\cite{RT}). We feel this is not a particularly good model for the 4-
dimensional case because 3-manifolds do not display any analog of
the problematic nilpotence phenomenon.

Following this idea we consider TQFT on 2-complexes. Formally this
is a functor from a category of 2-complex ``bordisms'' to a category of
modules over a ring. The objects in the bordism category are graphs
(1-complexes). A morphism $G_1\to G_2$ is an equivalence class of
2-complexes containing the disjoint union $G_1\sqcup G_2$. The
equivalence relation is Andrews-Curtis deformation through 2-
complexes, leaving $G_1\sqcup G_2$ fixed.

Denote the functor by $Z$, so a graph $G$ has an associated $R$-
module $Z(G)$, and a 2-complex bordism $X\:G_1\to G_2$ has an
associated homomorphism $Z_X\:Z(G_1)\to Z(G_2)$.  These take
disjoint unions to tensor products: $Z(G_1\sqcup G_2)\simeq
Z(G_1)\otimes_RZ(G_2)$, and similarly for morphisms.

The empty set is taken to $R$. To get invariants of a 2-complex
without specified subgraphs regard it as a bordism from the empty
graph to itself. This gives a homomorphism $Z_X\:R\to R$, or
equivalently an element $Z_X(1)\in R$.

We relate this back to the nilpotence problem. The class of TQFT we
consider has the property that the homomorphisms $Z_X$ are
unchanged by attaching  1-complexes to $X$. The 1-point union
$X\vee S^2$ is equivalent to  $S^2$ attached  by an arc to $X$. The
induced homomorphism is unchanged by deleting the arc. But
disjoint unions are taken to products, so
$$Z_{X\vee S^2}=Z_{X\sqcup S^2}=(Z_X)(Z_{S^2}). $$
If $Z_{S^2}$ is a unit in $R$ then we can recover $Z_{X}$ from
$Z_{X\vee S^2}$, and $Z$ cannot detect counterexamples to the
conjecture. At the other extreme if $Z_{X\vee S^2}=0$ then $Z$ has
the same sort of ``sudden death'' instability that the Donaldson
invariants do with respect to connected sum with $S^2\times S^2$.

Therefore we seek TQFT on 2-complexes for which  $Z_{S^2}=0$. In
the ones constructed from representations $Z_{S^2}=\sum(\rank
V_i)^2$, which is why we want this identity. It is hard to imagine
how such a TQFT could be nontrivial without detecting something
new, but at present there is no general proof of this. It would have to
be verified by calculations of examples.

\head 3. Representations\endhead
Fix a commutative ring $R$. The complex numbers is probably a good
choice: we will end up with units $\rank V_i$ and will want the sum
of their squares to vanish. Let $H$ be a Hopf algebra over $R$ which
is triangular in the sense of Drinfel'd \cite{Dr1}. We really do mean
triangular, and not quasitriangular.

Suppose $V$, $W$ are representations of $H$ (ie\. modules which are
finitely generated projective as $R$-modules). Then the coproduct in
$H$ gives a way to put an $H$-module structure on the product
$V\otimes_RW$. Denote this new representation by $V\sq W$. The
coassociativity required in a Hopf algebra makes this operation
associative. The triangular structure makes it commutative in the
sense there is a canonical isomorphism $\Psi\:V\sq W\to W\sq V$,
and the square of this is the identity.

We now formulate the problem: Find $H$ so that there are
representations $V_0, \dots, V_n$ satisfying
\roster\item there are $H$-morphisms $\lambda_i\:V_i\sq V_i\to R$
which are symmetric in the sense $\lambda_i=\lambda_i\Psi$, and
nondegenerate as bilinear forms;
\item each $V_i\sq V_j$ is isomorphic to a sum of copies of the
$V_*$;
\item the $V_i$ are irreducible and distinct; and
\item $\sum (\rank V_i)^2=0$.
\endroster
\subhead Remarks\endsubhead

\subsubhead {\rm (i)} Quasi-Hopf algebras\endsubsubhead These
conditions are much like the definition of a ``modular Hopf algebra''
used in \cite{RT}, \cite{Wa}, etc. This reflects a basic similarity in the
construction. The triangular Hopf algebra structure is used to give
the category of representations a symmetric monoidal (tensor)
category structure (\cite{Mc},  \cite{DM}). For this it is sufficient to
have a triangular quasi-Hopf algebra in the sense of \cite{D2}.

\subsubhead {\rm (ii)} Ranks\endsubsubhead
 The rank used in \therosteritem4 is the categorical one: the trace of
the identity homomorphism. More explicitly let $V^*$ denote the
dual $\hom_R(V,R)$ with the induced $H$ structure, then
representations are reflexive in the sense that the natural map
$i\:V\to V^{**}$ is an isomorphism. Evaluation is a morphism
$\ev\:V\sq V^*\to R$. Form the composition
$$R@>\ev^*>>V^*\sq V^{**}@>\id\sq i^{-1}>>V^*\sq V@>\Psi>>V\sq
V^*@>\ev>>R$$
then this is multiplication by some element of $R$. That element is
defined to be the rank of $V$.  It follows from \therosteritem1 and
\therosteritem3 that $\rank V_i$ is a unit.

\subsubhead {\rm (iii)} The functor $E$\endsubsubhead
Define a functor $E$ from $H$-representations to $R$-modules by
$$E(V)=\hom_H(R,V)/\{x\:R\to V\mid fx=0 \text{ for every }f\:V\to
R\}.$$
Roughly this is the trivial subrepresentation of $V$ modulo the
elements which cannot be detected by an invariant function  $V\to
1$. When $V$ has a nondegenerate form on it as in \therosteritem1
then $E(V\sq V)\simeq E(V\sq V^*)$ is essentially the
endomorphism ring of $V$.

\subsubhead {\rm (iv)} Irreducibility\endsubsubhead
The condition in \therosteritem3 is correct if $R$ is an algebraically
closed field (eg\. $\CC$). The general conditions are $E(V_i\sq
V_i)\simeq R$ as an algebra, and $E(V_i\sq V_j)=0$ for $i\neq j$.
Since $V_i$ is self-dual $E(V_i\sq V_i)$ is basically the
endomorphism ring, so the first condition is ``irreducibility.'' The
second condition asserts roughly that the $V_i$ are distinct.

\subsubhead {\rm (v)} Essential isomorphisms\endsubsubhead
The construction uses the functor $E$, so it is sufficent for the
conditions to hold ``on the $E$ level.'' Specifically we say two
morphisms $f,g\:V\to W$ are  {\it essentially equal\/} if
$E(f\sq\id)=E(g\sq\id)$ for any identity $\id\:U\to U$.
We caution that  it is not enough simply to assume $E(f)=E(g)$.

Thus \therosteritem1 can be weakened to ``$\lambda_i$ essentially
equal to $\lambda_i\Psi$,'' and for \therosteritem2 we need only an
essential isomorphism $\sum n_iV_i\to V_j\sq V_k$. The analog of
\therosteritem2 in \cite{RT} is only true up to essential isomorphism.

\subsubhead {\rm (vi)} Strategy\endsubsubhead The form of the
conditions suggests a strategy for finding examples somewhat like
the construction of Drenfel'd in \cite{D2}.

Note that the associativity isomorphism $U\sq(V\sq W)\simeq (U\sq
V)\sq W$ does not appear in the conditions; it only has to exist. The
commutativity isomorphism $\Psi\:U\sq V\simeq V\sq U$ appears
in the symmetry hypothesis in \therosteritem1 and the definition of
the rank in \therosteritem4. Perhaps one could begin with a Hopf
algebra and representations $V_i$ satisfying the first three
conditions. Then adjust the $R$-matrix in $H$, from which $\Psi$ is
defined, to make \therosteritem4 true. The forms in \therosteritem1
may no longer be symmetric with respect to the new $\Psi$, but it
seems likely that new ones can be found. Finally solve for a
``coassociativity'' structure as in \cite{D2} to get a triangular quasi-
Hopf algebra satisfying the conditions.

\head 4. A sketch of the construction\endhead
We actually get a {\it modular\/} TQFT in the sense of \cite{Q3},
generalizing a definition of G. Segal. This means we have relative
modules $Z(G,P)$ defined for a graph containing a finite set of points
$P$. Graphs can be glued together along points in the specified
subsets, and there are natural homomorphisms
$$Z(G_1,P_1\sqcup P_2)\otimes_R Z(G_2,P_2\sqcup P_3)\to
Z(G_1\cup_{P_2}G_2,P_1\sqcup P_3).\tag{1}$$
Note if $G_1$ is the unit interval and is attached along one end then
the union is equivalent to $G_2$. This gives a ring structure on
$Z(I,\partial I)$, and module
structures over this ring on $Z(G,P)$, one for each point in $P$.
Putting these together gives a ring structure on $Z(P_2\times
I,P_2\times\partial I)$ and a module
structure over this on $Z(G_1,P_1\sqcup P_2)$.

The modularity axiom then asserts that the homomorphism (1)
factors through an isomorphism on the tensor product over this ring,
 $$Z(G_1,P_1\sqcup P_2)\otimes_{Z(P_2\times I,P_2\times\partial
I)} Z(G_2,P_2\sqcup P_3)\to Z(G_1\cup_{P_2}G_2,P_1\sqcup
P_3).\tag{2}$$

We now begin the construction. Let $V=\sum V_i$ be the sum of the
representations given by the data. If $G$ is contractible  (a tree) then
we define
$$Z(G,P)=E(\sq^PV).\tag{3}$$
The right side of this is the functor $E$ of \S3 applied to a product of
copies of $V$, one copy for each point in $P$. In fact for this to make
sense we must choose orderings and associations to describe the $P$-
fold product as a sequence of 2-fold products. We obtain something
independent of the choice by using an inverse limit. More precisely
we define a little category whose objects are the possible orderings
and associations of $P$, and morphisms generated by changing one
association or changing the order of two adjacent objects. The
associativity and commutativity isomorphisms in the representation
category give a functor from this little category into the
representations. Compose with $E$ and take the inverse limit
(defined in \cite{Mc}).

It is at this point that the triangularity of the Hopf algebra is needed.
Up to appropriate equivalence we can think of the tree $G$ as the
cone on $P$. The symmetries of the cone must be reflected in
symmetries of $Z(G,P)$, so permuting points in $P$ must give an
action of the permutation group. In the 3-manifold constructions the
analogous step uses $G$ a punctured sphere and $P$ a union of
circles. Symmetries of a punctured sphere which permute the
boundary circles are given by the braid group rather than the
permutation group. To get an action of the braid group on $Z(G,P)$ it
is sufficient to take an inverse limit over a category whose
morphisms are ``braid equivalences'' of orderings and associations.
And to define a functor on this category one only needs a quasi-
triangular structure on the Hopf algebra, ie\. the isomorphisms
$\Psi\:U\sq V\to V\sq U$ need not have order two.

If $G=G_1\sqcup G_2\sqcup\cdots\sqcup G_n$ is a disjoint union of
trees, then define $Z(G,P)=\otimes_{i=1}^nZ(G_i,P \cap G_i)$.  This
agrees with the disjoint union property expected of a TQFT.

Now define some of the natural homomorphisms (1). Suppose we are
joining two contractible graphs  at a single point $p$. Then define the
homomorphism to be the composition
$$E(\sq^{P_1\sqcup p}V)\otimes E(\sq^{p\sqcup
P_2}V)@>>>E(\sq^{P_1\sqcup p\sqcup p\sqcup
P_2}V)@>>>E(\sq^{P_1\sqcup P_2}V)$$
where the first takes $x\otimes y$ to $x\sq y$ (recall these are
morphisms from $R$ into various representations), and the second
uses the symmetric form $\lambda$ to contract the  repeated copies
of $V$ associated to the two copies of $p$.

This defines the ring structures on $Z(I,\partial I)$, and as remarked
in \S3 it is more or less the endomorphism ring of $V$. We can think
of the actions of this ring on $Z(G,P)=E(\sq^PV)$ as induced from the
actions of the endomorphism ring on the components of $\sq^PV$.
The modularity condition (2) follows easily in these cases.

Now define $Z(G,P)$ for a general finite graph. Cut the graph at points
$Q$ to make all components contra

{\bf Note added by Greg Kuperberg on 2/18/01: Garbled material 
material has been removed from the TeX submission here.  A discontinuity
in the narrative remains.}

\bye